\documentstyle[12pt]{article}
\input epsf.sty
\def\ZZZ{{\hbox{ Z\kern-1.6mm Z}}}

\newcommand{\vt}{\vartheta}

\newcommand{\eps}{\epsilon}

\newcommand{\ppl}{\vec P_{\parallel L}}

\newcommand{\FF}{{\cal F}}

\newcommand{\OO}{{\cal O}}

\newcommand{\wt}{\widetilde}
\newcommand{\wh}{\widehat}

\newcommand{\NN}{{\cal N}}

\newcommand{\be}{\begin{equation}}
\newcommand{\ee}{\end{equation}}
\newcommand{\ben}{\begin{eqnarray}\displaystyle}
\newcommand{\een}{\end{eqnarray}}
\newcommand{\refb}[1]{(\ref{#1})}
\newcommand{\p}{\partial}
\newcommand{\sectiono}[1]{\section{#1}\setcounter{equation}{0}}

\def\one{{\hbox{ 1\kern-.8mm l}}}
\def\zero{{\hbox{ 0\kern-1.5mm 0}}}
 
\begin{document}
{}~
{}~
\hfill\vbox{\hbox{hep-th/0504005}
}\break
 
\vskip .6cm
\begin{center}
{\Large \bf 
Black Holes and
the Spectrum of Half-BPS States in N=4 
Supersymmetric String Theory
}

\end{center}

\vskip .6cm
\medskip

\vspace*{4.0ex}
 
\centerline{\large \rm
Ashoke Sen}
 
\vspace*{4.0ex}

\centerline{\large \it Harish-Chandra Research Institute}

\centerline{\large \it  Chhatnag Road, Jhusi,
Allahabad 211019, INDIA}
 
\centerline{E-mail: ashoke.sen@cern.ch,
sen@mri.ernet.in}
 
\vspace*{5.0ex}
 
\centerline{\bf Abstract} \bigskip

The entropy of a half-BPS black hole in $N=4$ supersymmetric 
heterotic string
compactification is independent of the details of the 
charge vector and is a
function only of the norm of the charge vector calculated using
the appropriate Lorenzian metric. Thus in order for this to agree
with the degeneracy of the elementary string states, the latter
must also be a function of the same invariant norm. We show that
this is true for generic
CHL compactifications to all orders in a power
series expansion in the inverse charges, but there are
exponentially suppressed
corrections which do depend on the details of the 
charge vector. This
is consistent with the hypothesis that the black hole entropy
reproduces the degeneracy of elementary string states to all
orders in a power
series expansion in the inverse charges, and helps us extend the
earlier conjectured relation between black hole entropy and
degeneracy of elementary string states to 
generic half-BPS electrically charged
states in generic $N=4$ supersymmetric heterotic
string compactification. Using this result we can also relate the
black hole entropy to the statistical entropy calculated using an
ensemble of elementary string states that contains all BPS states
along a fixed null line in the lattice of electric charges.

\vfill \eject
 
\baselineskip=18pt

\tableofcontents

\sectiono{Introduction and Summary} \label{sintro}

Recent
progress\cite{0409148,0410076,0411255,0411272,0501014,0502126,0502157}
towards relating the degeneracy of half-BPS states
in the spectrum of elementary string\cite{rdabh0,rdabh1}
to the entropy of the black
hole carrying the same charge quantum 
numbers\cite{9504147,9506200,9712150} opens up the
possibility of exploring 
the relationship
between the black hole entropy and the statistical entropy
in string theory in much
more detail than has been achieved so far. In carrying out a similar
analysis for systems involving
D-branes and other non-perturbative objects in string
theory\cite{9601029,9711053} one encounters the problem that while
it is relatively easy to compute the degeneracy of BPS states in the
limit of large charges, 
it is not always possible to make a more precise estimate that is valid
even for finite values of the various charges.
In contrast the degeneracy of elementary string states can be
calculated precisely in a wide variety of theories in which the
string world-sheet theory is described by a solvable conformal field
theory. This in turn should make it possible to carry out
a detailed comparison of this degeneracy
with the black hole entropy that goes
beyond the large charge limit.

However, in carrying out this program
we encounter a conceptual difficulty:
finite size corrections to entropy
(and other thermodynamic quantities) depend on the ensemble that we
use, and hence there is no unambiguous expression for the statistical
entropy even if we know the spectrum exactly.
This is due to the fact that in order to establish the
equivalence between different ensembles we need to replace the
ensemble average of any quantity by the value of the quantity
for the most probable configuration. This is a valid approximation
in the infinite size limit where the probability distribution is
sharply peaked around the most probable configuration, but is not
strictly valid for a finite size system.
Thus in order to carry out a comparison between the entropy of a black
hole to the degeneracy of elementary string states beyond the large
charge limit,
we first need to decide which ensemble we should use to compute the
statistical entropy of elementary string states. Since there is as yet
no fundamental principle that determines this, we can pose the question
in a slightly different way:
is there a statistical ensemble for
half-BPS elementary string states such that the entropy computed using
this ensemble agrees with the entropy of the black hole carrying the
same charge quantum numbers?

So far there has been two different
proposals for the choice of ensembles.
The first one uses an expression for the
black hole entropy computed without taking into account the
effect of holomorphic anomaly in the expression for the generalized
prepotential, and relates it to the statistical entropy
computed using an ensemble where we keep half of the charges
(`magnetic charges')
fixed
and sum over all possible values of the other charges after introducing
chemical potential conjugate to these 
charges\cite{0405146,0409148,0412139,0502157}. The second
proposal uses an expression for the black hole entropy computed after
taking into account the effect of holomorphic
anomaly\cite{9302103,9307158,9309140} following \cite{9906094},
and relates
it to the statistical entropy of an ensemble where a single $T$-duality
invariant combination of the charges is allowed to 
vary\cite{0411255,0502126}.\footnote{A different version of this ensemble
for studying entropy of quarter BPS black holes have been suggested
in \cite{0412287}.}
The relationship between
these two proposals is not completely clear at present.

In this paper we continue our study of the
second proposal.
We find that in order to generalize this
proposal to generic
half BPS states in a generic $\NN=4$ supersymmetric heterotic
string compactifications, the spectrum of half-BPS states in these
theories must satisfy some consistency conditions. In particular,
the degeneracy should be expressible as a continuous function of the
charges and should not jump by `large' amount as we move from one
point on the charge lattice to a neighbouring point, even if the
two charges arise from different twisted sectors of an
orbifold theory.
We then go ahead
and analyze the spectrum of half-BPS states in these theories to 
show that these consistency conditions are satisfied, 
and furthermore that the spectrum agrees with the
proposed formula for relating the black hole entropy
to the statistical
entropy of half-BPS elementary string states up to exponentially
suppressed correction terms. We also give an
alternative interpretation of the ensemble used in computing the
statistical entropy. In this interpretation the ensemble 
includes all states whose
charges lie along a fixed null line in the lattice of physical electric
charges.

Since part of the paper is somewhat technical, we shall now
summarize the main results in some more detail.
Let us consider an $\NN=4$ supersymmetric heterotic string
compactification\cite{CHL,CP,9507027,9507050,9508144,9508154}
with $(22-k)$ matter multiplets. This theory
has a rank $(28-k)$ gauge group, and a generic eletrically charged
state is labelled by a  $(28-k)$ dimensional
electric
charge vector.
There is a natural T-duality invariant inner product
of signature $(22-k, 6)$ in the $(28-k)$ dimensional lattice of
electric charges.  Let us consider
a black hole solution in this theory carrying
electric
charge vector $Q$, and let $N=Q^2/2$ where
$Q^2$ denotes the $SO(22-k,6)$
invariant norm of this charge vector. It turns out that $S_{BH}$
is a function of $N$ alone and does not depend on the details
of the vector $Q$. 
We now define $\FF_{BH}(\mu)$ as the Legendre transform
of $S_{BH}(N)$:
\be \label{es1}
\FF_{BH}(\mu) = S_{BH}(N) - \mu N\, ,
\ee
where $N$ is the solution of the equation
\be \label{es2}
{\p S_{BH}(N)\over \p N} - \mu = 0\, .
\ee
The proposed relation between the 
degeneracy $d(Q)$
of elementary string states carrying charge
$Q$ and $\FF_{BH}(\mu)$ is\footnote{As usual, in computing $d(Q)$ we
sum over all angular momentum states.}
\be \label{es3}
d(Q) \simeq {1\over 2\pi i} \, e^{-C_0}
\, \int_{
\mu_0 - i a}^{\mu_0 + ia} \, d\mu
\, e^{\FF_{BH}(\mu)+\mu N}\, ,
\ee
where $a$ is a sufficiently small but fixed positive 
number but is otherwise arbitrary,
$e^{-C_0}$ is an overall normalization
constant whose value we are unable to determine due to
an uncertainty in computing the additive constant
in the expression for
$S_{BH}$, and $\mu_0$ is the saddle point of the
integrand on the positive
real axis, corresponding to the solution of the equation:
\be \label{ex}
\FF'_{BH}(\mu_0) + N = 0\, , \qquad Im(\mu_0)=0, \qquad 
Re(\mu_0) >0\, .
\ee
The $\simeq$ in \refb{es3} denotes
equality up to error terms
which are suppressed relative to the leading term by powers of
$e^{-\pi\sqrt N}$. As long as $a$ is sufficiently small the
contribution to \refb{es3} is dominated by the saddle point
of the integrand given in \refb{ex},
and the result of doing the
integration is independent of $a$
up to terms that are suppressed by
powers of $e^{-\pi\sqrt N}$. If we simply replace the integral by
the value of the integrand at the saddle point, the right hand
side of \refb{es3} reduces to the exponential of the inverse
Legendre
transform of $\FF_{BH}(\mu)$
up to a multiplicative constant.
This gives us back the usual leading order relation
$\ln d(Q)=S_{BH}$ up to an additive
constant. However \refb{es3}
is a refined version of this proposal
which is supposed to reproduce
$\ln d(Q)$ to all orders in a power series expansion in $1/N$.
One of the consequences of the proposal
given in \refb{es1}-\refb{es3}
is that if we ignore corrections to $d(Q)$
which are suppressed by powers
of $e^{-\pi\sqrt N}$ then $d(Q)$ must be a function $d_N$
only of the combination
$N=Q^2/2$ and should not depend on the details of the charge vector
$Q$.

Explicit computation of the black hole entropy shows that
the part of $\FF_{BH}(\mu)$ that contributes
to \refb{es3} up to terms
suppressed by powers of $e^{-\pi\sqrt N}$ has the
form
\be \label{eks1}
\FF_{BH}(\mu) \simeq {4\pi^2\over \mu} + {24-k\over 2} 
\ln{\mu\over 2\pi}
+ C \, ,
\ee
where
$C$ is a constant whose value is not known at 
present. Substituting
this into \refb{es3}, \refb{ex} we get
\be \label{eks2}
d(Q) \simeq {e^{C-C_0} \over 2\pi i} 
\, \int_{\mu_0 - i a}^{\mu_0 + ia} \, d\mu \, 
\exp\left( {4\pi^2\over \mu} + {24-k\over 2} 
\ln{\mu\over 2\pi} + \mu N \right)\, ,
\ee
and
\be \label{emueq}
-{4\pi^2\over \mu_0^2} +{24-k\over 2\mu_0} + N = 0 \, .
\ee 
Note that $\mu_0$ calculated from \refb{emueq}
is a function of $N$, but we shall avoid
displaying this dependence explicitly in order to avoid
cluttering up the various formul\ae.

The proposed relation \refb{es3} differ from that in 
\cite{0405146,0409148,0412139,0502157} in
two important ways. First of all the black hole
entropy is computed using the S-duality invariant
one particle irreducible effective action that includes explicit 
non-holomorphic corrections to the generalized prepotential
instead of the Wilsonian
effective action where such non-holomorphic corrections are absent. 
Second, the Laplace transform in \refb{es3}
is taken with respect to a single variable $\mu$ conjugate to
the combination $N=Q^2/2$ instead of the chemical potentials
conjugate to all the electric charges.

In section \ref{srev} we shall review the result for 
$S_{BH}(N)$
that leads to \refb{eks1}. We also
give various different (but equivalent) versions of the
proposal \refb{es3}. 
For example, we show that \refb{es3} may be reexpressed as
\be \label{ex1}
\FF_{BH}(\mu) \simeq \FF(\mu) + C_0 \, ,
\ee
where
\be \label{eexp2}
\exp\left(\FF(\mu)\right) 
\equiv {1\over M}\, \sum_{N} d_{N} e^{-\mu N}\, .
\ee
Here $d_N$ is the degeneracy of elementary string states with
$Q^2=2N$ and $M$ is an integer that counts the number of allowed
values of $N$ per unit interval.
$\simeq$ in \refb{ex1} denotes equality up to terms
suppressed by powers of $e^{-\pi^2/\mu}$.
The sum over $N$ in \refb{eexp2} runs over all allowed values of $N$
in the theory. In computing $d_N$ on the right hand side of
\refb{eexp2} we must pick a specific representative charge vector
$Q$ satisfying $Q^2=2N$ and identify $d_N$
as the degeneracy
of these states. Different representatives differ from each other 
by terms suppressed by powers of $e^{-\pi\sqrt N}$ which introduces
an uncertainty in $\FF(\mu)$ that is suppressed by powers
of $e^{-\pi^2/\mu}$. This does not affect our analysis of the
proposed relation \refb{ex1} since this relation is expected
to be valid in the small $\mu$ limit
up to exponentially suppessed terms.

Taking the inverse Legendre transform on
both sides, we may also express \refb{ex1} as:
\be \label{eexp1}
S_{BH}(N) \simeq \wt S_{stat}(N) + C_0 \, ,
\ee
where 
$\wt S_{stat}(N)$ is the `statistical entropy'
of half-BPS elementary string states defined through the relation:
\be \label{eexp1a}
\wt S_{stat}(N') = \FF(\mu) + \mu N'\, , 
\qquad  {\p\FF(\mu)\over \p\mu}+N'=0\, .
\ee
$\simeq$ in \refb{eexp1} denotes equality up to error terms which are
suppressed by powers of $\exp(-\pi\sqrt N)$.

In section \ref{sver} we shall verify the proposal
\refb{es3} relating black hole entropy and degeneracy of elementary string 
states
by
computing the latter in $\NN=4$ 
supersymmetric heterotic string compactifications. 
In particular, verification of \refb{es3}
requires proving
that up to correction
terms suppressed by powers of $e^{-\pi\sqrt N}$, the result for
$d(Q)$ depends only on
the combination $Q^2$ and not on the details of the charge
vector $Q$.
The functional dependence on $Q^2$ (including the overall normalization) 
must be the same irrespective
of whether the state arises in the untwisted sector or one of the
twisted sectors of the theory. 

In the form given in eqs.\refb{ex1}-\refb{eexp1},
the proposed relations 
are equivalent to identifying the black hole entropy to the entropy
of half-BPS elementary string states in an ensemble that contains
BPS states of different charges, counting all the states with a
given $Q^2$ value $2N$
as one state. $\mu$ denotes the chemical potential conjugate to the
variable $N=Q^2/2$. This looks a bit odd since $Q^2$ is a
quadratic function
of the charges and hence is not an additive quantum number.
A more natural choice of an ensemble would be one in which
we keep some charges fixed, introduce a chemical potential conjugate to
the other charges, and sum over all possible values of the other charges.
We demonstrate in section \ref{sreint} that such an interpretation is
possible for the partition function defined in \refb{eexp2}. We take two
fixed charge vectors $Q_0$ and $s_0$ on the lattice of physical electric
charges with $s_0$ being null, and consider an 
ensemble that contains all states
of the form $Q_0+ n s_0$ with $n\in Z$.
If $\beta$ denotes the chemical
potential conjugate to the integer $n$, then the partition function
$\exp\left(\wh\FF(Q_0, s_0, \beta)\right)$ of this ensemble has a 
simple relation to
the partition function $\exp\left(\FF(\mu)\right)$ introduced in 
\refb{eexp2}. Thus the proposed relation \refb{eexp1}
may also be regarded as the statement
of equivalence between the black hole
entropy and the statistical entropy associated with an ensemble
of elementary
string states whose charges lie along a null line in the lattice.

\sectiono{Entropy of Half BPS Black Holes and its 
Relation
to the Degeneracy of Elementary String States} \label{srev}

Let us consider an $\NN=4$ heterotic string compactification
with $(22-k)$ massless matter
multiplets\cite{CHL,CP,9507027,9507050,9508144,9508154}.
This theory has altogether
$(28-k)$ U(1) gauge fields, of which 6 are graviphoton fields
arising out of the right-moving currents on the world-sheet and $(22-k)$
are part of the matter multiplets arising out of the left-moving
currents on the world-sheet. We shall refer to these gauge fields as
right-handed and left-handed gauge fields respectively.
We consider a  half-BPS black
hole carrying $(28-k)$ dimensional electric charge vector $Q$.
There
is a natural Lorenzian metric of signature $(22-k,6)$ in the
vector space of the charges. We define
\be \label{e1}
N = Q^2 /2 \, ,
\ee
where $Q^2$ is 
the norm of $Q$
measured with this metric. At any given point in the moduli space of the
theory the $(28-k)$ dimensional vector space has a natural decomposition
into a direct sum of a $(22-k)$ dimensional vector space of left-handed
electric charges and a six dimensional vector space of right-handed
electric charges. The latter charges couple to the graviphoton fields.
If we denote by $\vec Q_L$ and $\vec Q_R$ the left and the right-handed
components of the charges, then
\be \label{eq2}
Q^2 = (\vec Q_R^2 - \vec Q_L^2) \, , \qquad \to \qquad N = {1\over 2}
\, (\vec Q_R^2 - \vec Q_L^2) \, .
\ee

The entropy of this black hole vanishes in 
the supergravity approximation.
However higher derivative corrections become
important near the horizon since
the curvature and other field strengths
become large in this region. There
is a general scaling argument that shows
that for large $Q^2$ the entropy
of the black hole, after taking into
account the higher derivative
corrections, goes as $a\sqrt{N}$ for
some constant $a$\cite{9504147,9712150}. More recently, 
refs.\cite{0409148,0410076,0411255,0411272,0501014,0502126,0502157}
computed the entropy of these black holes
by taking into account a specific class
of higher derivative terms in the action,
following
earlier work on supersymmetric attractor
mechanism\cite{9508072,9602111,9602136} and
the effect of higher derivative terms on the black hole
entropy\cite{9602060,9603191,
9812082,9904005,9906094,9910179,0007195,0009234,0012232}.
The higher derivative terms which were used in this calculation
represent corrections to
the generalized prepotential of the theory and are of the
form\cite{rzwiebach,9610237,9708062,9906094}
\be \label{eb6}
\int d^4 x \, \sqrt{\det g} \left[
\phi(S, \bar S) \, R^-_{\mu\nu\rho\sigma} 
R^{-\mu\nu\rho\sigma} + c.c.\right] + \ldots \, ,
\ee
where $g_{\mu\nu}$, $R^\pm_{\mu\nu\rho\sigma}$ and $S$ 
denote respectively 
the canonical
metric, the self-dual and anti-self-dual components of the
Riemann tensor and the complex scalar field whose real and
imaginary parts are given by the exponential of the dilaton field 
and the axion field respectively. The function\footnote{We
shall be considering those heterotic string compactifications
which admit dual type II description so that we can compute
$\phi(S,\bar S)$ by working in this dual
description\cite{9610237,9708062,0502126}.
A wide class of such models were constructed in
\cite{9508144,9508154}.}
\be \label{eb7a}
\phi(S,\bar S)=g(S) -{K\over 128\pi^2} \, \ln(S+\bar S)
\ee 
is the sum of a piece $g(S)$ that is holomorphic
in $S$ and a piece proportional to $\ln(S+\bar S)$
that is a function of both $S$ 
and $\bar S$. For large $S$ the function $g(S)$ has the form
\be \label{egs}
g(S) = {S\over 16\pi} + \OO\left(e^{-2\pi S}\right) \, .
\ee
Furthermore, the combination $\wh h(S)$, defined as
\be \label{egsa}
\wh h(S)\equiv \p_S g(S) +{K\over 32\pi^2} \, \p_S \ln 
\eta(e^{-2\pi S})\, , \qquad \eta(q)\equiv q^{1/24} 
\prod_{n=1}^\infty (1 - q^n) \, ,
\ee
transforms as a modular form of weight two under the S-duality
group of the theory\cite{0502126}.
The constant $K$ in \refb{eb7a}, \refb{egsa}
is given by\cite{9708062,0502126}
\be \label{ekvalue}
K = 24-k\, ,
\ee
and represents the effect of holomorphic 
anomaly\cite{9302103,9307158,9309140}.
Finally, $\ldots$ in \refb{eb6} denotes various other terms
which are required for the supersymmetric completion. 

In the presence of the term given in \refb{eb6}, the black hole
entropy $S_{BH}$ is given by the formula\cite{9906094,0502126}:
\be \label{eb4aa}
S_{BH} = {\pi N \over S_0} + 
64\pi^2 \, g(S_0) - {K\over 2} \, \ln (2S_0) + C\, ,
\ee
where $S_0$, the value of the field $S$ at the horizon, is
determined from the equation:
\be \label{eb5aa}
-{\pi\, N\over S_0^2} + 64 \pi^2 g'(S_0) -{K\over 2S_0} \simeq 0\, ,
\ee
and $C$ is a constant whose value is not known at present.
Note that $S_{BH}$ is a function only of
the combination $N$ defined in \refb{eq2} and does not depend
on the details of the charge vector $Q$.
In arriving at eqs.\refb{eb4aa}, \refb{eb5aa} one needs to 
take into account the correction
to the Bekenstein-Hawking formula for the 
entropy due to higher derivative
terms in the action\cite{9307038,9312023,9403028,9502009}.
It should be mentioned however that this formula has been derived from
first principles only in the case $K=0$ where the non-holomorphic
contribution to $\phi(S, \bar S)$ is absent. It is usually difficult
to supersymmetrize the non-holomorphic terms, and \cite{9906094}
guessed this formula for toroidally compactified heterotic string theory
using the requirement of S-duality invariance. Eqs.\refb{eb4aa} and
\refb{eb5aa} are generalizations of this formula for general $\NN=4$
supersymmetric compactification\cite{0502126}. The constant $C$ in
\refb{eb4aa} is not determined by the
requirement of duality invariance. We should keep in  mind however
that for a given theory this is a fixed constant, and is independent of
the charges carried by the black hole.

We now define $\FF_{BH}(\mu)$ as
the Legendre transform of the function $S_{BH}(N)$ with respect to
the variable $N$. Then $S_{BH}$ is the inverse Legendre transform of
$\FF(\mu)$:
\be \label{en3}
S_{BH}(N) = \FF_{BH}(\mu) + \mu N\, ,
\ee
with $\mu$ determined from the formula
\be \label{en4}
{\p \FF_{BH}\over \p\mu} + N = 0\, .
\ee
\refb{eb4aa}, \refb{eb5aa} are identical to \refb{en3},
\refb{en4} provided we make the identification
\be \label{en1}
\mu = {\pi\over S_0}\, ,
\ee
\ben \label{en2}
\FF_{BH}(\mu) &=& 64\pi^2 \, g\left(\pi/\mu\right) - 
{K\over 2} \, \ln 
(2\pi/\mu) 
+ C \nonumber \\
&=& {4\pi^2 \over \mu} - {24-k\over 2} \, \ln 
\left({2\pi\over\mu}\right) 
+ C + \OO\left( e^{-2\pi^2/\mu}\right)
\een
In arriving at the aecond line of this equation we have used
\refb{egs}, \refb{ekvalue}.
$e^{\FF_{BH}(\mu)}$ will be called the black hole partition function.

We propose the following relation between $\FF_{BH}(\mu)$ and  
the degeneracy $d(Q)$ of
elementary string states carrying electric charge $Q$:
\be \label{en30n}
d(Q) \simeq  {1\over 2\pi i} \, {e^{-C_0}}
\, \int_{\mu_0-ia}^{\mu_0+ia}
e^{ \FF_{BH}(\mu) + \mu N}\, , \qquad N={1\over 2} \, Q^2\, ,
\ee
where $\simeq$ denotes equality up to error terms which
are suppressed
by powers of $e^{-\pi\sqrt N}$ relative to the leading term,
$\mu_0$ is given by the solution of the equation
\be \label{emueqearly}
\FF'_{BH}(\mu_0)+ N =0, \qquad Im(\mu_0)=0, \qquad Re(\mu_0)> 0\, ,
\ee
$a$ is a small
positive constant about which we shall say more in the
next paragraph
and $C_0$ is a constant whose value will be determined
in eq.\refb{eident}. Note that according to this proposal
the degeneracy of the elementary string states should be a function
of the combination $N=Q^2/2$ only, and should not depend on the
details of the charge vector $Q$. In section \ref{sver}
we shall verify that this is true up to exponentially suppressed
correction terms.

In order to understand the role of the limits of integration on
the right hand side of \refb{en30n}, we need to analyze this integral in 
some
detail.
{}From eq.\refb{en2}
it follows that $\mu_0$
given in eq.\refb{emueqa} is a saddle point of the integrand in
\refb{en30n} up to exponentially suppressed corrections.
Up to power corrections $\mu_0\simeq 2\pi/\sqrt N$, and 
the value of the
integrand near this saddle point has a factor of
$e^{4\pi\sqrt{N}}$,
with a factor of $e^{2\pi\sqrt N}$ coming from the
$e^{\FF_{BH}(\mu_0)}$ term and another factor of $e^{2\pi\sqrt N}$
coming from the $e^{\mu_0 N}$ term.
As mentioned below eq.\refb{egsa}, the function $g(\pi/\mu)$ 
appearing in the
expression \refb{en2}
for $\FF_{BH}(\mu)$ has modular properties, and as a 
result
there are additional divergences of the
integrand near the points of the form $\mu=i\nu$, 
$\nu=2\pi p/q$ for integer
$p$, $q$. Near these points ${\FF_{BH}(\mu)}
\sim {4 \pi^2 c_{p,q}/(\mu - i\nu)}$
for some constant $c_{p,q}$ that depends
on the fraction $p/q$ and typically is smaller for larger $q$.
Since the integration
contour has been chosen to have $Re(\mu)=\mu_0$, the maximum
contribution to the integrand
from a singularity at $\mu=i\nu$ comes when the integration
contour passes through the point $\mu=\mu_0+i\nu$, and at this
point the integrand has a factor of order 
$e^{2\pi\sqrt{N}(1+c_{p,q})
+ i\nu \, N}$. As long as $c_{p,q}$ is smaller than 1, this
contribution is exponentially suppressed compared to the
contribution coming from the saddle point near $\mu_0$. 
We shall choose the constant $a$
appearing
in the integration range of $\mu$ to be
sufficiently small so that for all points of the form
$2\pi i p/q$ in the range
$(-ia, ia)$ the coefficient $c_{p,q}$ is smaller than 1.
In this case
the dominant contribution to the integral comes from
a region close to $\mu_0$, and hence
the dependence of the integral on the constant $a$ will be
exponentially suppressed as long as $a$ satisfies the critera
given above.
Also in this case we can ignore the contributions to
$\FF_{BH}(\mu)$ which are suppressed by powers of
$e^{-\pi^2/\mu}$,
since after integration this will produce corrections that are
suppressed by powers of $e^{-\pi^2/\mu_0}\sim e^{-\pi\sqrt N/2}$.
Eqs.\refb{en2} now allows us to restate the proposal
\refb{en30n}, \refb{emueqearly} as
\be \label{ednbh}
d(Q) \simeq {1\over 2\pi i} \, {e^{-C_0}}
\, \int_{\mu_0-ia}^{\mu_0+ia}
\exp\left(\mu N + {4\pi^2 \over \mu} - {24-k\over 2} \, \ln 
\left({2\pi\over\mu}\right) 
+ C \right)\, ,
\ee
where $\mu_0$ is the solution of the equation
\be \label{emueqa}
-{4\pi^2\over \mu_0^2} +{24-k\over 2\mu_0} + N = 0 \, , 
\qquad \mu_0> 0\, .
\ee

We shall now reformulate this relation in terms of a relation
between the black hole
partition function $e^{\FF_{BH}(\mu)}$ and the
partition function of elementary string states, defined as
\be \label{e2a}
e^{\FF(\mu)} = {1\over M}\, \sum_{N} d_N \, e^{-\mu N}\, ,
\ee
where the sum over $N$ in \refb{e2a}
runs over all the allowed values of $N$ in the theory, and
$M$ is an integer that counts the number of
allowed values of $N$ per unit interval. In order to define the
quantity $d_N$ appearing on the
right hand side of \refb{e2a} we need to pick a representative
charge vector $Q$ so that $Q^2/2=N$ and identify $d_N$ as the
degeneracy $d(Q)$ of states carrying charge $Q$. As mentioned
earlier, the leading contribution of order $e^{4\pi\sqrt{N}}$
to $d_N$ is independent of the choice of the representative $Q$
that we choose, but there are
corrections suppressed by powers
of $e^{-\pi\sqrt N}$ which do depend on this representative. By
analyzing the sum in \refb{e2a} by saddle point method one can
easily see that a contribution to
$d_N$ that grows as $\exp(4\pi c\sqrt 
N)$ will give rise to a contribution to $e^{\FF(\mu)}$ of order 
$\exp(4\pi^2 c^2/\mu)$ for small $\mu$. Thus an 
uncertainty in $d_N$ that 
is suppressed by powers of $\exp(-\pi\sqrt N)$ will induce
an uncertainty in the definition
of $\FF(\mu)$ that is suppressed by powers of $e^{-\pi^2/\mu}$. 
This will not affect
our analysis, since the proposed relation \refb{en4a} between
$\FF_{BH}(\mu)$ and $\FF(\mu)$ is expected to be valid only
up to exponentially
suppressed correction terms.

In order to analyze the consequence
of the proposed form \refb{ednbh} for 
$d_N$ for the partition function
$e^{\FF(\mu)}$, it will be convenient to 
work with a more general set of
partition functions defined as follows. 
For any integer $J$, let $S_J$ denote 
the set of allowed values of $N$ in 
the interval $-1\le N < J-1$. Since there are $M$ possible 
values of $N$ in unit interval, the 
set $S_J$ will contain $JM$ elements.
For any $\alpha\in S_J$, we define
\be \label{en15}
e^{\FF^{(J)}_\alpha(\mu)} \equiv J \, 
\sum_{L=0}^\infty d_{JL+\alpha} 
e^{-\mu(J L+\alpha)}\, . 
\ee
As in the definition of $\FF(\mu)$,
$\FF^{(J)}_\alpha(\mu)$ also
suffers from an uncertainty that is suppressed by powers of
$e^{-\pi^2/\mu}$. Using \refb{e2a}, \refb{en15} we get
\be \label{en16}
e^{\FF(\mu)} = {1\over JM}\sum_{\alpha\in S_J} 
e^{\FF^{(J)}_\alpha(\mu)}\, .
\ee

We shall now substitute \refb{ednbh}
into \refb{en15} to
come up with a proposal for the form of $\FF^{(J)}_\alpha(\mu)$:
\be \label{euni1}
e^{\FF^{(J)}_\alpha(\mu)} \simeq {J\over 2\pi i} \,
e^{-C_0+C}
\, 
\sum_{L=0}^\infty e^{-\mu(JL+\alpha)} \int_{\mu_0
-ia}^{\mu_0+ia}
\, d\mu' \, 
e^{4\pi^2/\mu' +\mu' (JL+\alpha)} 
\left({\mu'\over 2\pi}\right)^{12-{1\over 2} k} \, , 
\ee
where $\simeq$ denotes equality up to
terms suppressed by powers of $e^{-\pi^2/\mu}$ compared to the
leading term. Note that in arriving at \refb{euni1} we have
used the proposed form of $d_N$ in terms of black hole entropy
that is
correct up to fractional
errors involving powers of $e^{-\pi\sqrt N}$,
but this is okay since
these induce errors in the expression
for $\FF^{(J)}_\alpha(\mu)$ involving
powers of $e^{-\pi^2/\mu}$.
In \refb{euni1} $\mu_0$ is the solution of eq.\refb{emueqa}
with $N$ replaced by $JL+\alpha$ and hence depends on $JL$.
We shall now deform the $\mu'$ integration
contour so that it runs parallel
to the real axis from
$\mu_0-i a$ to a point close to the point
$-i a$ on the imaginary axis, then runs 
till a point close to $i a$ by grazing the imaginary axis,
and finally returns back to $\mu_0+i a$ along a line parallel to the
real axis. 
For the first and the third component of the contour, the 
$e^{4\pi^2/\mu'}$ part remains finite, and hence the contribution to the 
$\mu'$ integral from these contours are 
bounded by a term of order $e^{\mu_0(JL+\alpha)}\sim e^{2\pi 
\sqrt{JL+\alpha}}$. After performing the sum over $L$ for these 
terms we get contributions  bounded by a term 
of order $e^{\pi^2/\mu}$. The middle component of the contour that grazes 
the 
imaginary $\mu'$ axis is independent of $L$. Furthermore for any point 
$\mu'$ on 
this contour
$Re(\mu-\mu')>0$
for any finite $\mu$. Hence we can
perform the sum over $L$ explicitly and write the contribution as
\be \label{euni2}
{J\over 2\pi i} \,
e^{-C_0+C} \, 
e^{-(\mu-\mu')\alpha} \int_{0^+-ia}^{0^++ia}
\, d\mu' \, {1\over 1 - e^{-J(\mu-\mu')}} \, 
e^{4\pi^2/\mu'}
\left({\mu'\over 2\pi}\right)^{12-{1\over 2} k} \, , 
\ee
We can now evaluate the integral by deforming the 
$\mu'$ integration contour to a 
semicircle of radius $a$ in the $Re(\mu')>0$ region. 
During this process we pick up the residue at the pole at 
$\mu'=\mu$. This contribution is given by
\be \label{euni3}
e^{-C_0+C} \, 
e^{4\pi^2/\mu} 
\left({\mu\over 2\pi}\right)^{12-{1\over 2} k} \, .
\ee
The left-over 
contour integral along the semi-circle $|\mu'|=a$
does not have any exponentially large contribution, and hence
this contribution is exponentially suppressed 
compared to \refb{euni3}. Furthermore, the contribution from the first and 
the third components of the contour, which were seen to be bounded by 
terms of order 
$e^{\pi^2/\mu}$, are also exponentially small compared to \refb{euni3}. 
Thus \refb{euni3} gives the complete contribution to \refb{euni1}
up to exponentially suppressed terms, and we get
\be \label{esa1}
\FF^{(J)}_\alpha(\mu) \simeq 
{4\pi^2\over \mu} +\left( 12-{1\over 2} k\right)
\ln {\mu\over 2\pi} \, +C - C_0 \simeq \FF_{BH}(\mu) - C_0\, .
\ee
\refb{esa1} implies that for small $\mu$,
$\FF^{(J)}_\alpha(\mu)$ is independent of $J$ and
$\alpha$ up to correction terms
involving powers of $e^{-\pi^2/\mu}$. 
This information, together eq.\refb{en16}, gives
\be \label{enn1}
\FF^{(J)}_\alpha(\mu) \simeq \FF(\mu)\, .
\ee
This in turn allows us to restate \refb{esa1} 
as\footnote{This is the
way the conjecture was stated earlier in refs.\cite{0411255,0502126}.
There we worked with a special class of theories
where the compactification manifold has an $S^1$
factor
and chose the representative $d_N$ in eq.\refb{en15} from
a special class of
elementary string states which carry $w$ units of
winding charge and $n$ units of 
momentum
along this circle
without carrying any other charges. For these states $Q^2=2nw$, and
one can check explicitly that $d_N$ depends only on the combination
$nw$ and not individually on $n$ and $w$.}
\be \label{en4a}
\FF(\mu) \simeq \FF_{BH}(\mu) - C_0 \, .
\ee

Eqs.\refb{esa1} and \refb{en4a} are different ways of restating the
proposed relation \refb{en30n} between black hole entropy and degeneracy 
of elementary string states. 
By taking the inverse Legendre transform of both sides of
eq.\refb{en4a} we may state the proposal in yet another
form:
\be \label{ere1}
\wt S_{stat}(N) \simeq  S_{BH}(N) - C_0 \, ,
\ee
where $\wt S_{stat}(N)$ is the inverse
Legendre transform of $\FF(\mu)$:
\be \label{ere2}
\wt S_{stat}(N) = \FF(\mu) + \mu N, \qquad 
{\p \FF(\mu)\over \p\mu}+N=0\, .
\ee
$\simeq$ in eq.\refb{ere1} represents equality up to error terms involving
powers of $e^{-\pi\sqrt N}$.

\sectiono{Counting Degeneracy of BPS States in CHL Compactification}
\label{sver}

In this section we shall generalize the analysis of 
refs.\cite{0502126,0502157} to
compute the degeneracy of half-BPS elementary
string states in a general class of CHL 
models\cite{CHL,CP,9507027,9507050,9508144,9508154} and verify
its proposed relation to the black hole entropy
as given in eqs.\refb{en30n}, \refb{ednbh}.
These models are
obtained by beginning with heterotic string compactified on $T^6$
and then modding out the theory by the action of an abelian group
$G$. 
In particular, refs.\cite{9508144,9508154}
constructed a wide class of such
models which admit dual type II description by beginning with
heterotic string compactification on a six dimensional torus
of the form $T^4\times S^1\times \wt S^1$,
and then modding out the
theory by a $Z_m\times Z_n$ group where $Z_m$ ($Z_n$) acts on 
$S^1$ ($\wt S^1$) by a shift of order $m$ ($n$) and also acts as a
$Z_m$ ($Z_n$) symmetry transformation on the rest
of the conformal field theory involving the coordinates of $T^4$
and the other 16 left-moving world-sheet scalars associated with
$E_8\times E_8$ gauge group.
The action of a typical element $g$ of the orbifold group can be
regarded as a combination of a shift $a_g$ and
a rotation $R_g$ on the
signature $(22,6)$ Narain lattice $\Gamma$
associated with
the toroidal compactification\cite{narain,nsw}:
\be \label{ech1}
P \to R_g P + a_g \, , \qquad P\in 
\Gamma\, .
\ee
The rotation part also acts on the oscillators.
The set of $R_g$'s for $g\in G$ form a group that describes the
rotational part of $G$. We shall call this group $R_G$.
In order to preserve
$\NN=4$ supersymmetry we need to ensure that $R_G$ acts
trivially on the right handed world-sheet
fields. If we assume that
the
full group $R_G$ leaves $(22-k)$ of the 22 left-moving directions
invariant,
then for any given $g$, $R_g$
can be characterized by $k/2$ rotation angles
$2\pi\phi_1(g),\ldots 2\pi\phi_{k/2}(g)$.\footnote{For a particular
element $g$ some of the $\phi_j(g)$'s may vanish, but for each $j$,
$\phi_j(g)\ne 0$ for at least one $g\in G$.}
We shall
denote by $X^j$ 
$(1\le j\le {k\over 2})$ the complex world-sheet scalars
labelling the planes
of rotation, so that the effect of rotation
on the corresponding left-handed oscillators $\alpha^j_{-n}$
is represented by
\be \label{ech1a}
\alpha^j_{-n}\to e^{2\pi i \phi_j(g)}
\alpha^j_{-n}\, .
\ee
The corresponding complex conjugate oscillator transforms with a phase
$e^{-2\pi i \phi_j(g)}$. Thus without any loss of generality we can restrict
$\phi_j(g)$'s to be in the range
\be \label{erange}
0\le \phi_j(g) \le {1\over 2}\, .
\ee

For future reference it will be useful to set up some
notations and definitions at this stage.
Let $V$ denote the $(22+6)$ dimensional
vector space in which the lattice $\Gamma$ is embedded. For
a given group element $g$, we denote by $V_\perp(g)$ the subspace
of $V$ that is left invariant by $R_g$, and by
$V_\parallel(g)$
its orthogonal subspace. Thus the planes of rotation associated with
a given group element $g$ are
orthogonal to $V_\perp(g)$ and lie along
$V_\parallel(g)$. We also define
\be \label{ech2}
V_\perp = \bigcap_{g\in G} V_\perp(g), \qquad V_\parallel=
\bigcup_{g\in G} V_\parallel(g)\, .
\ee
In others words $V_\perp$ is the subspace of $V$ that is left invariant
by the entire group $G$, and $V_\parallel$ is its orthogonal
subspace. The total dimension of $V_\perp$ is $(28-k)$. This
is the number of U(1) gauge fields in the resulting theory.

Finally, let us define
\be \label{ech3}
\Lambda_\perp(g) = \Gamma\bigcap V_\perp(g)\, , \qquad
\Lambda_\perp = \Gamma\bigcap V_\perp = \bigcap_g \Lambda_\perp(g)\, ,
\ee
and
\be \label{ech4}
\Lambda_\parallel(g) = \Gamma\bigcap V_\parallel(g)\, ,
\qquad \Lambda_\parallel = \Gamma\bigcap V_\parallel\, .
\ee

Let us now begin analyzing the spectrum
of half-BPS
elementary string states. We first consider untwisted sector states.
Before the orbifold projection a generic BPS state is obtained
by keeping the right-moving sector of the state at the lowest
$L_0$ eigenvalue allowed  by GSO projection, and considering
arbitrary excitations on the left-moving sector. If we denote
by $(\vec P_L, \vec P_R)$ the left and right components of the
momentum vector, and by $N_L$ the total level of left-handed
oscillator excitations, then the level matching condition gives
\be \label{ech6}
N_L-1 + {1\over 2} (\vec P_L^2 - \vec P_R^2) = 0\, .
\ee
While in the original theory all the components of $P$ act as
sources for electric fields, in the orbifold theory only the
components of $P$ along $V_\perp$ act as sources for electric
fields. We shall denote by $Q=(\vec Q_L, \vec Q_R)$ the projection
of $P$ along $V_\perp$ and by $P_\parallel=(\ppl, 0)$ the projection
of $P$ along $V_\parallel$. 
The requirement of $\NN=4$
supersymmetry forces $V_\parallel$ to be directed fully along the
left-handed component of the lattice. As a result $P_\parallel$
has no right-handed component and $\vec P_R=\vec Q_R$. 
Eq.\refb{ech6} may now be rewritten as:
\be \label{ech7}
N_L - 1 + {1\over 2} \, \ppl^2 = N\, ,
\ee
where
\be \label{ech8}
N \equiv {1\over 2} (\vec Q_R^2 - \vec Q_L^2)\, .
\ee
If two vectors $P$ and $P'$ in $\Lambda$ correspond to the same charge
vector $Q$, then $P-P'=(\vec P_{\parallel L} - 
\vec P'_{\parallel L} , 0)$. Thus $\vec P_{\parallel L} - 
\vec P'_{\parallel L} \in \Lambda_\parallel$. This shows that for
a given $Q$, the allowed values of $\ppl$ are of the form
\be \label{eform}
\ppl = \vec K(Q) + \vec p \, ,
\qquad \vec p \in \Lambda_\parallel\, ,
\ee
where $\vec K(Q)$ is a fixed vector in $V_\parallel$
lying within the unit cell
of $\Lambda_\parallel$.

We want to count the number of $G$ invariant BPS states $d(Q)$ for a 
fixed $Q$. This is best done by taking the trace of 
$\sum_{g\in G} g$
over all
states carrying the given electric charge $Q$ and then dividing
the answer by the total number of elements $n_G$ of the group. 
Clearly the contribution to $Tr(g)$ will come from only those
$\ppl$ which are invariant under $R_g$. This requires
\be \label{ergpk}
\vec p + \vec K(Q) \in V_\perp(g)\, .
\ee
Acting on the Fock vacuum carrying such a momentum $P$,
the
group element $g$ produces a phase 
\be \label{ephase}
e^{2\pi i a_g\cdot P} = e^{2\pi i a_g \cdot Q} e^{-2\pi i \vec a_{gL}
\cdot (\vec p + \vec K(Q))}\, .
\ee 
If we
denote by $d^{(osc)}(N_L, g)$ the number of ways we can construct total
oscillator level $N_L$ out of the twenty four left-moving oscillators,
weighted by the action of the group element $g$ on that combination, then
we may express the total number of BPS states carrying charge $Q$ as
\ben \label{ech12}
d(Q) &=& 
{16\over n_G} \sum_{g\in G}
\sum_{N_L=0}^\infty
d^{(osc)}(N_L, g) e^{2\pi i a_g \cdot Q} 
e^{-2\pi i \vec a_{gL} \cdot \vec K(Q)} \nonumber \\
&& \sum_{\vec p \in \Lambda_\parallel \atop 
\vec p + \vec K(Q) \in V_\perp(g)}
 \, e^{-2\pi i \vec a_{gL}
\cdot \vec p}\, \delta_{N_L -1 + {1\over 2}
(\vec p + \vec K(Q))^2, N}
\een
The factor of 16 in this equation counts the number of states in a
single BPS supermultiplet.

In a sector twisted by a group element $g'$, the oscillators
of $X^j$
are fractionally moded with modes
$\alpha^j_{-n+\phi_j(g')}$. In this
case the momenta $P$ carried by the state lie in the vector
space $V_\perp(g')$ and are of the
form\cite{asymmetric}
\be \label{ech13}
P=\wt a_{g'} + \lambda, \qquad \lambda \in  \Lambda_\perp(g')^*\, ,
\ee
where $\wt a_{g'}$ is the projection of the vector $a_{g'}$
into $V_\perp(g')$,
and 
$\Lambda_\perp(g')^*$ is the lattice
dual to $\Lambda_\perp(g')$.
If we denote by $N_L$ the total level of the left-handed oscillators,
and by $(\vec P_L, \vec P_R)$ the left and the right handed
components of the charge vector as before, 
the level matching condition
reads
\be \label{eleveltw}
N_L-1+{1\over 2} (\vec P_L^2 - \vec P_R^2) + {1\over 2}
\sum_{j=1}^{k/2}
\phi_j(g') 
(1 - \phi_j(g')) = 0, 
\ee
where the last term on the left hand side of this equation accounts
for the $\bar L_0$ eigenvalue of the ground
state of the twisted sector.
As before the electric charge vector
$Q=(\vec Q_L, \vec Q_R)$ is given by the projection of
$P$ into $V_\perp$. For a generic CHL compactification
of the type described in \cite{9508144,9508154}, all states
carrying a given charge $Q$ arises
from a single twisted sector since the projection of
$a_{g'}$ into
$V_\perp$ are different for different $g'$. We shall denote
by $g'_Q$ the twist associated with the charge vector $Q$.
We also denote by $P_\parallel=(\ppl,0)$ the projection
of $P$ into $V_\parallel$. 
Then using \refb{ech13} and
an argument identical to the one that
led to eq.\refb{eform} we can show that
$\ppl$ must have the form:
\be \label{ech15}
\ppl = \vec K(Q) + \vec p, \qquad \vec p \in \Lambda_1(g'_Q)\, ,
\ee
where 
\be \label{edefl1}
\Lambda_1(g') = \Lambda_\perp(g')^*\bigcap V_\parallel\, ,
\ee
and $\vec K(Q)$ is some vector in 
$V_{\parallel}\bigcap V_\perp(g'_Q)$
within the unit cell of the lattice 
$\Lambda_\perp(g'_Q)^*\bigcap 
V_\parallel$. Eq.\refb{eleveltw} now
takes the form:
\be \label{eleveltwa}
N_L-1+{1\over 2} (\vec p + \vec K(Q))^2 + 
{1\over 2} \, \sum_{j=1}^{k/2}
\phi_j(g'_Q) 
(1 - \phi_j(g'_Q)) = N\, , \qquad N\equiv 
{1\over 2}(\vec Q_R^2-\vec Q_L^2)\, .
\ee

As in the case of untwisted sector states, 
we want to count the number of $G$ invariant BPS states $d(Q)$ for a 
fixed $Q$ in a sector twisted by some fixed
group element $g'_Q$.
Again this is done by taking the trace of $\sum_{g\in G}g$ over all
states carrying the given electric charge $Q$ and then dividing
the answer by the total number of elements $n_G$ of the group. 
Nonvanishing contribution to the trace will come from only those
$\ppl$ which are invariant under $R_g$.
Thus we must have
\be \label{exx1}
\vec p + \vec K(Q) \in V_\perp(g)\, .
\ee
Let us denote
by $d^{(vac)}(g')$ the
degeneracy of the ground state in the sector 
twisted by $g'$, -- this is the generalization of the number of
fixed points for a symmetric orbifold. 
We shall label these vacua by the index $r$ running
from 1 to $d^{(vac)}(g')$. 
Acting on the $r$th Fock vacuum carrying momentum $P$,
the
group element $g$ produces a phase
\be \label{ephasea}
e^{i\chi_r(g'_Q,g)}\,
e^{2\pi i a_g\cdot P} \, ,
\ee 
where we have allowed for the possibility that acting on the
$r$-th vacuum in the sector
twisted by $g'$, the action of the group element $g$ may produce a
momentum independent phase $e^{i\chi_r(g',g)}$. We note however that
if $g$ is the identity element then this phase must be trivial:
\be \label{echigp}
e^{i\chi_r(g',1)}=1 \, .
\ee
If we now
denote by $d^{(osc)}(N_L,g', g)$ the number of ways we can construct total
oscillator level $N_L$ out of the twenty four left-moving oscillators in
the sector twisted by $g'$, 
weighted by the action of the group element $g$ on that combination, 
and 
then
we may write
\ben \label{ech12a}
d(Q) &=& {16\over n_G} \, \sum_{r=1}^{d^{(vac)}(g'_Q)}
\, \sum_{g\in G
} e^{i\chi_r(g'_Q,g)}\,
\sum_{N_L=0}^\infty
d^{(osc)}(N_L, g'_Q,g) e^{2\pi i a_g \cdot Q} 
e^{-2\pi i \vec a_{gL} \cdot \vec K(Q)} \nonumber \\
&& \sum_{\vec p
\in \Lambda_1(g'_Q) 
\atop \vec p +\vec K(Q)\in V_\perp(g)
}  \, e^{-2\pi i \vec a_{gL}
\cdot \vec p}\, \delta_{N_L -1 +{1\over 2} (\vec p + \vec 
K(Q))^2
+{1\over 2}
\sum_{j=1}^{k/2} \phi_j(g'_Q) (1 - \phi_j(g'_Q)), N}\, ,
\een
where we have used the fact that
for a momentum of the type given in \refb{ech15},
\be \label{epdef}
a_g\cdot P = 
a_g \cdot Q - a_{gL}
\cdot (\vec p + \vec K(Q)) \, .
\ee
Note that eq.\refb{ech12} can be regarded as a special case of 
eq.\refb{ech12a} by setting $g'_Q=1$ in the latter equation. Thus in
once we compute $d(Q)$ using \refb{ech12a}, we do not need to compute
\refb{ech12} separately.

In order to evaluate the right hand side
of \refb{ech12a}, we shall rewrite this equation in a
slightly different way. We  write
\be \label{eabs1}
d(Q) = F(Q,N) \, , \qquad N\equiv 
{1\over 2}(\vec Q_L^2-\vec Q_R^2)\, ,
\ee
where
\ben \label{eabs2}
F(Q,\wh N) &\equiv& {16\over n_G}  \, \sum_{r=1}^{d^{(vac)}(g'_Q)}
\,
\sum_{g\in G
} e^{i\chi_r(g'_Q,g)}\,
\sum_{N_L=0}^\infty
d^{(osc)}(N_L, g'_Q,g) e^{2\pi i a_g \cdot Q} 
e^{-2\pi i \vec a_{gL} \cdot \vec K(Q)} \nonumber \\
&& \sum_{\vec p
\in \Lambda_1(g'_Q)\atop \vec p +\vec K(Q)\in V_\perp(g)
}  \, e^{-2\pi i \vec a_{gL}
\cdot \vec p}\, \delta_{N_L -1 +{1\over 2} (\vec p + \vec 
K(Q))^2
+{1\over 2} \sum_{j=1}^{k/2} \phi_j(g'_Q) (1 - \phi_j(g'_Q)), \wh N}\, .
\een
Note that \refb{eabs2} we have regarded 
$\wh N$ as an independent variable
not related to $Q$. This allows 
us to introduce a `partition function':
\be \label{eabs3}
\wt F(Q, \mu)
= \sum_{\wh N} \, F(Q,\wh N) \, e^{-\mu \wh N} \, ,
\ee
where the sum over $\wh N$ runs over all values for which 
$F(Q,\wh N)$
is non-zero. By the left-right level matching condition, these
are of the form\footnote{This follows from the fact that $F(Q,\wh N)$
counts the number of states in the conformal field theory which
carry charge $Q$, have their right-handed oscillator excitations
at the minimal level allowed by GSO projection, and have
$(\bar L_0-L_0)$ eigenvalue $\wh N - {1\over 2} Q^2$. Since
the requirement of one loop modular invariance forces all states
in the CFT to carry integer $\bar L_0- L_0$ eigenvalue,
$\wh N - {1\over 2}Q^2$ must be an integer. Note however that
individual terms in the sum in 
\refb{eabs2} do not satisfy the $\wh N-{1\over 2}Q^2=$integer
condition. Only
after summing over $g$, which
corresponds to projecting onto $G$ invariant states, the unwanted
terms cancel.} 
\be \label{eabs4}
\wh N=N_0 + f(Q)\, ,
\ee
where $N_0$ is an integer and $f(Q)$ is a fixed number
between 0 and
1  which measures the fractional part of
$N\equiv{1\over 2}Q^2$.
Thus we also have the reverse relation:
\be \label{eabs5}
F(Q,\wh N) = {1\over 2\pi i} \int_{\eps-i\pi}^{\eps+i\pi}
\wt F(Q, \mu)\, e^{\mu \wh N}\, ,
\ee
where $\eps$ is any real positive number. 
We shall first
compute $\wt F(Q,\mu)$, and then use \refb{eabs5}
to compute $F(Q,\wh N)$. 

Using eqs.\refb{eabs2} and \refb{eabs3} we get
\ben \label{eabs6}
\wt F(Q,\mu) &=&
{16 \over n_G} 
\exp\left( \mu \left (1 -{1\over 2}
\sum_{j=1}^{k/2} \phi_j(g'_Q) (1 - \phi_j(g'_Q))
\right) \right) \nonumber \\
&& \sum_{r=1}^{d^{(vac)}(g'_Q)} \sum_{g\in G
} e^{i\chi_r(g'_Q,g)}\, 
e^{2\pi i a_g \cdot Q} 
e^{-2\pi i \vec a_{gL} \cdot \vec K(Q)} 
\wt F^{(osc)}(g'_Q,
g, \mu) \wt F^{(lat)}(Q,
g, \mu) \, , \nonumber \\
\een
where
\be \label{eabs7}
\wt F^{(osc)}(g',
g, \mu) =  \sum_{N_L=0}^\infty
d^{(osc)}(N_L, g',g) e^{-\mu N_L} \, ,
\ee
and
\be \label{eabs8}
\wt F^{(lat)}(Q,
g, \mu) =
\sum_{\vec p
\in \Lambda_1(g'_Q)\atop \vec p +\vec K(Q)\in V_\perp(g)
}  \, e^{-2\pi i \vec a_{gL}
\cdot \vec p}\, \exp\left(-{1\over 2}\mu (\vec p + \vec 
K(Q))^2 \right)
\, .
\ee

In order to compute $\wt F^{(osc)}(g',g, \mu)$ we note that
under the action of the group element $g$, the oscillator
$\alpha^j_{-n+\phi_j(g')}$ picks
up a phase of $e^{2\pi i \phi_j(g)}$. For a given group
element $g'$, let us denote by the $A(g')$ the subset
of the $k/2$ 
indices $j$ for which the $\phi_j(g')$ are non-zero and by
$B(g')$ the set complementary to $A(g')$ in the
set $(1,2,\ldots k/2)$.\footnote{Thus the $x^j$'s for $j\in A(g')$
and their complex conjugate coordinates span the 
vector space $V_\parallel(g')$.}
In this case
\be \label{eorder}
\hbox{order}(A_{g'}) = {1\over 2} \hbox{dim} V_\parallel(g')
\equiv {1\over 2} \, k_{g'}, \qquad \hbox{order} (B_{g'}) = 
{1\over 2} (k-k_{g'})\, .
\ee
Let us define
\be \label{eabs9}
\tau = {i\mu\over 2\pi}\, , \qquad q = e^{-\mu}=e^{2\pi i\tau}\, .
\ee
Then $\wt F^{(osc)}(g',g, \mu)$ is given by
\ben \label{ech5}
\wt F^{(osc)}(g',g, \mu)
&=& \left(\prod_{n=1}^\infty {1\over 1 - q^n}\right)^{24-k} 
\prod_{j\in B(g')}  \left(\prod_{n=1}^\infty {1\over 1 - 
e^{2\pi i \phi_j(g)} \, q^n} {1\over 1 - 
e^{-2\pi i \phi_j(g)} \, q^n}
\right) \nonumber \\
&& \prod_{j\in A(g')}   
\left(\prod_{n=1}^\infty {1\over 1 - 
e^{2\pi i \phi_j(g)} \, q^{n-\phi_j(g')}} \prod_{n=0}^\infty
{1\over 1 - 
e^{-2\pi i \phi_j(g)} \, q^{n+\phi_j(g')}} 
\right) \, . \nonumber \\
\een
Using the Jacobi $\vt$ function
\be \label{ein1}
\vt_1(z|\tau) = 2 q^{1/12} \, \eta(q) \, \sin(\pi z) \prod_{n=1}^\infty
(1 - q^n e^{2\pi i z}) (1 - q^n e^{-2\pi i z})\, ,
\ee
where
\be \label{ein2}
\eta(q) = q^{1/24} \, \prod_{n=1}^\infty (1 - q^n) \, , 
\ee
we can rewrite \refb{ech5} as
\be \label{ein3}
\wt F^{(osc)}(g',g, \mu)
= q\, \left(\eta(q)\right)^{{3k\over 2} - 24} 
\prod_{j\in B(g')} {2 \sin(\pi\phi_j(g)) \over 
\vt_1(\phi_j(g)|\tau)}
\, \prod_{j\in A(g')} {e^{i\pi (\phi_j(g) - \tau\phi_j(g'))}
\over i  \vt_1(\phi_j(g)-\tau\phi_j(g')|\tau)}\, .
\ee
If some of the $\phi_j(g)$'s for $j\in B_{g'}$ vanish, then we
should replace the corresponding term in the product by its
limit as $\phi_j(g)\to 0$:
\be \label{elimit}
\lim_{\phi_j(g)\to 0} {2 \sin(\pi\phi_j(g)) \over 
\vt_1(\phi_j(g)|\tau)}
= (\eta(q))^{-3}\, .
\ee

We are interested in the behaviour of $\wt
F^{(osc)}(g',g,\mu)$ for
small $\mu$, \i.e. for small $\tau$, since we shall see later that
up to exponentially suppressed corrections the contribution to
\refb{eabs5} comes from a small region around the origin. 
Using the modular transformation
properties of $\eta(\tau)$ and $\vt_1(z|\tau)$ it can be seen that
in this limit, if $z$ remains fixed with $0\le Re(z)\le {1\over 2}$, 
\be \label{ein4a}
\eta(q) \simeq e^{-{\pi^2\over 6\mu}} 
\sqrt{2\pi\over \mu}\, , 
\ee
\qquad
\be \label{ein4b}
\vt_1(z|\tau) \simeq e^{-{\pi^2\over 2\mu}} \,
e^{2\pi^2 z(1-z)/\mu} \, \sqrt{2\pi\over \mu} \, ,
\ee
where $\simeq$ denotes equality up to terms which are suppressed by
powers of $e^{-\pi^2/\mu}$. Using \refb{ein3}-\refb{ein4b} 
we see that in the $\mu\to 0$ limit the ratio
\be \label{eratio}
\wt F^{(osc)}(g', g, \mu) /\wt F^{(osc)}(g',1,\mu) \, ,
\ee
is exponentially small for any $g\ne 1$ due to the non-vanishing
$\phi_j(g)$'s.
As a result $\wt
F^{(osc)}(g'_Q,1,\mu)$, for which all the $\phi_j(g)$'s
vanish, is exponentially large compared to all other 
$\wt F^{(osc)}(g'_Q,g,\mu)$ appearing in \refb{eabs6}.
It is easy to see that
the $\wt F^{(lat)}(Q,g,\mu)$ factor cannot compensate for this
suppression, -- indeed from
\refb{eabs8} it follows that
$\wt F^{(lat)}(Q,1,\mu)$ is greater than or equal to $\wt F^{(lat)}
(Q,g,\mu)$ for any $g$. Thus the sum over $g$ in \refb{eabs6}
can be replaced by a single term corresponding to
$g=1$ if we are
willing to ignore corrections involving powers of $e^{-\pi^2/\mu}$.

This shows that in order to compute $\wt F(Q,\mu)$ up to
exponentially suppressed contributions, we only need to
evaluate $\wt F^{(osc)}(g'_Q,1,\mu)$ and $\wt F^{(lat)}(Q,1,\mu)$.
This simplifies the analysis enormously since all the 
$\phi_j(g)$'s vanish in \refb{ein3}, and hence 
all the terms in the set $B_{g'}$ are now
replaced by the right hand side of \refb{elimit}. Using \refb{eorder}
this gives
\be \label{ein5}
\wt F^{(osc)}(g',1,\mu) =
q\, \left(\eta(q)\right)^{{3\over 2}k_{g'} - 24} 
\, \prod_{j\in A(g')} {e^{-i\pi \tau\phi_j(g')}
\over i  \vt_1(-\tau\phi_j(g')|\tau)}\, .
\ee
In the $\tau\to 0$ limit 
\be \label{ein5a}
\vt_1(-\tau \phi_j(g')|\tau) \simeq - i \, \sqrt{2\pi\over \mu}\, 
e^{-{\pi^2\over 2\mu}} \, e^{{\mu\over 2} \phi_j(g')^2}\, 
2 \, \sin(\pi \phi_j(g'))\, 
\ee
up to exponentially suppressed terms.
Eqs.\refb{ein4a},
\refb{ein5} and \refb{ein5a} now give
\ben \label{ein6}
\wt F^{(osc)}(g',1,\mu) &\simeq& 
e^{4\pi^2/\mu} \exp\left( -\mu \left (1 - {1\over 2}
\sum_{j=1}^{k/2} \phi_j(g') (1 - \phi_j(g'))
\right) \right) \, \left( {\mu\over 2\pi}\right)^{12 - 
{1\over 2}k_{g'}} \nonumber \\
&& \, \prod_{j\in A_{g'}} 
{1\over 2\sin(\pi \phi_j(g'))} \, .
\een
In writing
the argument of the exponential in \refb{ein6} we have
replaced the sum over $j\in A_{g'}$ by the sum over all values of
$j$ in the range $(1, k/2)$, since $\phi_j(g')$ vanishes outside
the set $A_{g'}$ anyway.

Let us now turn to the analysis of $\wt F^{(lat)}(Q,1,\mu)$.
In this case $V_\perp(g=1)=V$ and the condition $\vec 
p+\vec K(Q)\in V_\perp(g)$ is trivially satisfied.
Thus eq.\refb{eabs8} 
simplifies to: 
\be \label{ein7}
\wt F^{(lat)}(Q,
1, \mu) =
\sum_{\vec p
\in \Lambda_1(g'_Q)} \exp\left(-{1\over 2}\mu (\vec p + \vec 
K(Q))^2 \right)
\, .
\ee
The dimension of the lattice $\Lambda_1(g')$ defined in \refb{edefl1}
is 
that of 
$V_\perp(g')\bigcap V_\parallel$. This counts the number of 
directions in the $k$ dimensional vector space $V_\parallel$ which
is left invariant under the element $g'$. Comparing this with
\refb{eorder} we see that
\be \label{ein9}
\hbox{dim} \Lambda_1(g') = k - k_{g'}\, .
\ee
With the help of eq.\refb{ein9} and Poisson resummation,
we may reexpress  \refb{ein7} as
\be \label{ein10}
\wt F^{(lat)}(Q,
1, \mu) = {1\over v_{\Lambda_1(g'_Q)}} \, 
\left({\mu\over 2\pi}\right)^{{1\over 2} (k_{g'_Q}-k)}\, 
\sum_{\vec q
\in \Lambda_1(g'_Q)^*} \exp\left(-{2\pi^2\over \mu}\,\vec q^2 
+2\pi i \vec q\cdot
\vec 
K(Q) \right)
\, ,
\ee
where for any lattice $\Lambda$, $v_\Lambda$ 
denotes the volume of the unit cell of the
lattice $\Lambda$ and $\Lambda^*$ denotes its dual lattice.
Thus up to exponentially suppressed contribution, we have
\be \label{ein11}
\wt F^{(lat)}(Q,
1, \mu) \simeq {1\over v_{\Lambda_1(g'_Q)}} \, 
\left({\mu\over 2\pi}\right)^{{1\over 2} (k_{g'_Q}-k)}\, .
\ee

Finally we need the value of
$d^{(vac)}(g')$, -- the degeneracy of the ground state of the
sector twisted by $g'$. This
was computed in ref.\cite{asymmetric} and is given by
\be \label{edvac}
d^{(vac)}(g') = {1\over 
v_{\Lambda_\perp(g')}}\, 
\prod_{j\in A(g')} (2\sin(\pi \phi_j(g'))\, .
\ee

We are now ready to compute $\wt F(Q,\mu)$. Restricting the sum
over $g$ in \refb{eabs6} to only over the identity element, 
and using eqs.\refb{echigp},
\refb{ein6}, \refb{ein11} and \refb{edvac} 
we get
\ben \label{etw2}
\wt F(Q,\mu) &\simeq& 16 \, {d^{(vac)}(g'_Q) \over n_G} 
\exp\left( \mu \left (1 - {1\over 2}
\sum_{j=1}^{k/2} \phi_j(g'_Q) (1 - \phi_j(g'_Q))
\right) \right) \nonumber \\
&& \, 
\times \, \wt F^{(osc)}(g'_Q,
1, \mu) \, \wt F^{(lat)}(Q,
1, \mu) \nonumber \\
&\simeq& {16\over n_G v_{\Lambda_1(g'_Q)}
v_{\Lambda_\perp(g'_Q)} }\, e^{4\pi^2/\mu} 
\left({\mu\over 2\pi}\right)^{12-{1\over 2} k} \, .
\een
{}From the definitions \refb{ech2} it
follows that $V_\perp(g')$ has an orthogonal decomposition:
\be \label{etw3}
V_\perp(g')=V_\perp \oplus (V_\perp(g')
\bigcap V_\parallel)\, .
\ee
Then, given any lattice $\Lambda\in V_\perp(g')$, we
have\cite{martinet}
\be \label{etw4}
v_{\Lambda\bigcap V_\perp} = v_{\Lambda} 
\, v_{\Lambda^*\bigcap V_\parallel}\, .
\ee
Choosing $\Lambda=\Lambda_\perp(g')$ and using 
\refb{edefl1} we now get
\be \label{etw5}
v_{\Lambda_\perp(g')\bigcap V_\perp} = 
v_{\Lambda_\perp(g')} \, v_{\Lambda_1(g')}\, .
\ee
However
\be \label{etw6}
\Lambda_\perp(g')\bigcap V_\perp = \Gamma\bigcap V_\perp(g')
\bigcap V_\perp
= \Gamma\bigcap V_\perp = \Lambda_\perp
\ee
is independent of $g'$. Using \refb{etw5}, \refb{etw6}
we may now reexpress \refb{etw2} as
\be \label{etw7}
\wt F(Q,\mu) \simeq {16\over n_G v_{\Lambda_\perp}
}\, e^{4\pi^2/\mu} 
\left({\mu\over 2\pi}\right)^{12-{1\over 2} k} \, .
\ee
Note that this expression 
(including its overall normalization) is
independent of the charge vector $Q$ irrespective of which twisted
sector it arises from. In the specific example of the $Z_2$ orbifold
model of \cite{CP} this feature can be seen explicitly in the results
of \cite{0502157}.

We can now try to compute $F(Q,\wh N)$ using \refb{eabs5}.
The choice of $\eps$ in \refb{eabs5} is arbitrary, but 
we shall find it convenient to take 
$\eps=\mu_0$ with $\mu_0$ given by the solution of
eq.\refb{emueqa} with $N$ replaced by $\wh N$. Thus
\be \label{eabs5a}
F(Q,\wh N) = {1\over 2\pi i} \int_{\mu_0-i\pi}^{\mu_0+i\pi}
\wt F(Q, \mu)\, e^{\mu \wh N}\, .
\ee
We might at this stage be tempted
to replace $\wt F(Q,\mu)$ in this expression by the right hand
side of \refb{etw7}.
However one needs to exercise a little more care,
since the relation \refb{etw7} holds only in
the region of small $\mu$, while the integration range over $\mu$
in \refb{eabs5a} extends over a finite range. Thus replacing 
$\wt F(Q, \mu)$ by
the right hand side of \refb{etw7} is possible only
if we can argue that the dominant contribution to the integral
in \refb{eabs5a} comes from a small
region around the origin.
{}From \refb{etw7}
it follows that $\mu_0$
given in eq.\refb{emueqa} is a saddle point of the integral
\refb{eabs5a} up to exponentially suppressed corrections.
Since up to power corrections $\mu_0\simeq 2\pi/\sqrt {\wh N}$, 
the value of the
integrand near this saddle point has a factor of
$e^{4\pi\sqrt{\wh N}}$,
with a factor of $e^{2\pi\sqrt {\wh N}}$ coming from the
${\wt F(Q,\mu_0)}$ term and another factor of $e^{2\pi\sqrt {\wh N}}$
coming from the $e^{\mu_0 \wh N}$ term.
{}From the modular properties of
$\wt F(Q,\mu)$ it
follows that there are additional divergences of the
integrand near the points of the form $\mu=i\nu$, 
$\nu=2\pi p/q$ for integer
$p$, $q$. Near these points ${\wt F(Q, \mu)}
\sim e^{4 \pi^2 c_{p,q}/(\mu - i\nu)}$ with $c_{p,q}<1$ as long as
$\mu$ is in the range $(-i\pi, i\pi)$.\footnote{Due
to the $\mu\to \mu+2\pi i$ periodicity,
at $\mu \simeq \mu_0 \pm 2\pi i$ 
we expect to get back a contribution of strength identical
to that near $\mu=\mu_0$,
but these points are outside the range of integration
in \refb{eabs5a}.}
Since the integration
contour has been chosen to have $Re(\mu)=\mu_0$, the maximum
contribution to the integrand
from a singularity at $\mu=i\nu$ comes when the integration
contour passes through the point $\mu=\mu_0+i\nu$, and at this
point the integrand has a factor of order $e^{2\pi\sqrt{\wh N}(1+c_{p,q})
+ i\nu \, \wh N}$. Since $c_{p,q}<1$, these  
contributions are
exponentially suppressed compared to the contribution
$\sim e^{4\pi\sqrt{\wh N}}$  from the
saddle point at $\mu_0$.
Thus
we see that up to exponentially suppressed terms,
the contribution to the integral comes from
a region close to $\mu_0$, and hence
in
the integral appearing in \refb{eabs5a}
we can replace $\wt F(Q,\mu)$ by the right hand side of
\refb{etw7}, and change the range of integration
to be from $\mu_0 - i a$ to 
$\mu_0 + i a$ where $a$ is some small
but fixed positive number. This gives
\be \label{etw11}
F(Q, \wh N) \simeq {1\over 2\pi i} \,
{16\over n_G v_{\Lambda_\perp}
}\, \int_{\mu_0 - ia}^{\mu_0+ia}\, d\mu \, 
e^{4\pi^2/\mu +\mu \wh N} 
\left({\mu\over 2\pi}\right)^{12-{1\over 2} k} \, .
\ee
Using \refb{eabs1} we now get
\be \label{euniversal}
d(Q) \simeq {1\over 2\pi i} \,
{16\over n_G v_{\Lambda_\perp}
}\, \int_{\mu_0 - ia}^{\mu_0+ia}\, d\mu \, 
e^{4\pi^2/\mu +\mu N} 
\left({\mu\over 2\pi}\right)^{12-{1\over 2} k} \, , \qquad
N\equiv {1\over 2}(\vec Q_R^2-\vec Q_L^2)\, .
\ee
This is in precise agreement with the proposed relation 
\refb{ednbh} provided we make the
identification
\be \label{eident}
{e^{-C_0+C}} = {16\over n_G v_{\Lambda_\perp}
}\, .
\ee
By the general arguments outlined in section \ref{srev}
this also establishes the relations
\refb{esa1} and \refb{en4a}
involving the partition functions, and the relation \refb{ere1}
involving the entropy.

\sectiono{A Reinterpretation of the Partition Function $\FF(\mu)$}
\label{sreint}

The definition of the partition function $e^{\FF(\mu)}$ corresponds to
choosing an ensemble where we introduce a chemical potential 
$\mu$ conjugate
to the combination $N=Q^2/2$. This is somewhat strange
since $N$ involves
the square of the charge vector,
and is not additive. It would seem more natural
to choose an ensemble where we keep some of the charges fixed and 
sum over all possible values of the other charges after introducing
a chemical potential conjugate to these charges.\footnote{For example
the ensembles used in \cite{0405146,0409148,0502157} is of this type.}
In this section we shall
show that due to the Lorentzian signature of the Narain lattice, and the 
universality of the expression for $d(Q)$, the
partition function $e^{\FF(\mu)}$ can also be reinterpreted in this
way.

Let us consider a fixed vector $Q_0$ in $V_\perp$
in the lattice of
physical charges, and let $s_0$ denote another fixed vector
in $V_\perp$ which is also in the lattice of physical charges,
and which furthermore is null. Then $Q_0+n s_0$ for any integer $n$
represents a physical charge, and for this state
\be \label{esb1}
N = {1\over 2} (Q_0+ns_0)^2 = {1\over 2} (Q_0)^2 + 
n Q_0\cdot s_0\, .
\ee
As long as $Q_0\cdot s_0\ne 0$,  we can 
choose $Q_0\cdot s_0$ to be positive without any loss of generality.
We now introduce an ensemble where we sum over all charges of the
form $Q_0 + n s_0$ for fixed $Q_0$ and $s_0$ by introducing a chemical
potential $\beta$ conjugate to the variable $n$:
\be \label{esb2}
\exp\left(\wh\FF(Q_0,s_0,
\beta)\right) = \sum_{n} d(Q_0+n s_0) e^{-\beta n}\, .
\ee
Using \refb{esb1} this reduces to
\be \label{esb3}
\exp\left(\wh\FF(Q_0,s_0,\beta)\right) 
\simeq \sum_n d_{{1\over 2}(Q_0)^2 + n Q_0\cdot s_0} e^{-\beta n}
\, ,
\ee
where $d_N$ as usual denotes the
universal formula for the degeneracy of 
half-BPS elementary
string states with $Q^2=2N$.
If $Q_0\cdot s_0=p/q$ for relatively prime integers $p$ and $q$, then we 
express $n$ as $jq + l$ with $0\le l\le q-1$, $j\in Z$, 
and rewrite the sum over $n$ in \refb{esb3} as
\ben \label{eaa1}
\exp\left(\wh\FF(Q_0,s_0,\beta)\right)
&\simeq& \sum_{l=0}^{q-1} \, \sum_{j\in Z} d_{{1\over 2}(Q_0)^2 + 
(jq+l){p\over 
q}} e^{-\beta (jq+l)} \nonumber \\
&=& e^{{1\over 2} \beta {q} Q_0^2/p} \, 
\sum_{l=0}^{q-1} \sum_{j\in 
Z} d_{{1\over 2}(Q_0)^2 + l{p\over q} + jp} e^{-({1\over 2}(Q_0)^2  
+ l{p\over q} + jp ) \beta {q/p}} \, .
\een
Using the definition \refb{en15} of $\FF^{(J)}_\alpha(\mu)$ and 
eq.\refb{enn1} this may be rewritten as\footnote{If the subscript
${1\over 2}(Q_0)^2 +l{p\over q}$ of $\FF^{(p)}$ lies outside the range
$(-1, p-1)$ then we need to bring it within this range by adding
appropriate integral multiples of $p$.}
\ben \label{eaa2}
\exp\left(\wh\FF(Q_0,s_0,\beta)\right)
&\simeq& e^{{1\over 2} \beta {q} Q_0^2/p} \sum_{l=0}^{q-1} {1\over p} 
\, \exp\left({\FF^{(p)}_{{1\over 2}(Q_0)^2 + l{p\over q}}(\beta 
q/p)}\right) \nonumber \\
&\simeq& e^{{1\over 2} \beta {q}  Q_0^2/p} \, {q\over p} \, 
e^{\FF(\beta 
q/p)} \nonumber \\
&=& {1\over Q_0\cdot s_0} \, e^ {\beta Q_0^2 \over 2 Q_0\cdot s_0} \, 
e^{\FF(\beta/ Q_0\cdot s_0)} \, .
\een
This gives
\be \label{esb4}
\wh\FF(Q_0,s_0,\beta) \simeq {\beta Q_0^2 \over 2 Q_0\cdot s_0} +
\FF(\beta/Q_0\cdot s_0) - \ln ( Q_0\cdot s_0)\, .
\ee
This gives a simple relation between the partition function 
$e^{\FF(\mu)}$
that we have used and the partition function
$e^{\wh\FF(Q_0,s_0,\beta)}$ defined in \refb{esb2}.
In particular if we
define $\wh S_{stat}(Q_0,s_0,n)$ as the Legendre transform of
$\wh\FF(Q_0,s_0,\beta)$:
\be \label{ekk1}
\wh S_{stat}(Q_0,s_0,n) = 
\wh \FF(Q_0,s_0,\beta) + \beta n, \qquad 
{\p \wh\FF(Q_0,s_0,\beta)\over
\p \beta} + n  = 0\, ,
\ee
then \refb{esb4} implies that that $\wh S_{stat}(Q_0,s_0,n)$
is related to the statistical
entropy $\wt S_{stat}(N)$ defined in eq.\refb{ere2} by the simple
relation:
\be \label{ekk2}
\wh S_{stat}(Q_0,s_0,n) \simeq \wt S_{stat}(N={1\over 2}(Q_0)^2 
+ n Q_0\cdot s_0) -
\ln (Q_0\cdot s_0)\, .
\ee
Thus the two entropies are essentially the same up
to an additive factor.
This additive factor reflects the difference in step size used in
defining the two ensembles. Using \refb{ekk2} we can now rewrite the 
conjectured relation \refb{ere1} as
\be \label{en}
\wh S_{stat}(Q_0,s_0,n) \simeq S_{BH}(N={1\over 2}(Q_0)^2
+ n Q_0\cdot s_0) - C_0 - \ln (Q_0\cdot s_0) \, .
\ee

\sectiono{Index or Absolute Degeneracy?} \label{sindex}

For $\NN=4$ supersymmetric string theories one can define an index
$\Omega_4$ which vanishes for a non-BPS state but is non-zero for
half-BPS states\cite{9708062}, and it has been suggested that the
computation of the black hole entropy by keeping only the corrections
to the generalized
prepotential and ignoring other higher derivative corrections
might lead to this index rather than the absolute
degeneracy\cite{0502157}.
Up to an overall normalization factor, this
index counts the number of short multiplets of the supersymmetry
algebra weighted by $(-1)^F$, where $F$ represents the space-time
fermion number of the highest $J_3$ state of the supermultiplet,
and
$J_3$ is
the third component of the angular momentum of the state.
In order to compute this index for fundamental string states
it is more convenient to express
$(-1)^F$ as $(-1)^{F_L} (-1)^{F_R}$, with $F_L$ and $F_R$
denoting the contribution to the space-time fermion number
from the left
and the right-handed sector of the world-sheet respectively. 
For any string compactification, if we consider half-BPS states
which involve excitations of the left-handed oscillators on the 
world-sheet, then each BPS multiplet can be regarded as a tensor
product
of a single BPS multiplet representing the ground state of the
right-handed oscillators and an arbitrary state involving the
left-handed oscillators.
$(-1)^{F_R}$ for the highest $J_3$ state of
such a BPS state is always $1$. Thus the computation of $\Omega_4$
involves computing the trace of $(-1)^{F_L}$ over the BPS states.

For heterotic string compactification, the left-moving sector does not
contribute to the fermion number. Hence each BPS state contributes 1
to $Tr(-1)^{F_L}$ and $\Omega_4$ is simply proportional to the absolute
number of BPS states\cite{0502157}. 
Thus in all the formul\ae\ we have given
in the earlier sections of this paper, {\it e.g.} eq.\refb{e2a},
we can replace the degeneracy $d_N$ by the index $(\Omega_4)_{N}$ up to 
an overall multiplicative factor, and the
analysis in the heterotic string theory cannot distinguish these two
prescriptions.

The situation however
is different in type II string theory. Let us consider for
example type II string theory compactified on a torus $T^6$, and consider 
a state for which the right-handed oscillators are in the lowest $L_0$
eigenvalue state consistent with GSO projection. In this case  at any
given level there are equal
number of states in the left-hand sector with $(-1)^{F_L}=1$
and $(-1)^{F_L}=-1$.
As a result the index $\Omega_4$ vanishes.
This is encouraging since it is known that for
type II
superstring theory on a torus, inclusion of higher derivative corrections
to the generalized prepotential 
does not produce a finite area for the black hole representing the 
fundamental type II string. Thus if in our formul\ae\ we replace the
degeneracy $d_N$ by the index $(\Omega_4)_{N}$, then the results agree
trivially since the leading contribution to both side of various
formul\ae\ vanish.

This however is not the end of the story. Consider for example type IIA
string theory on $T^4\times \wt S^1\times S^1$ and take the quotient of
this theory by a $Z_2$ group that acts as $(-1)^{F_L}$ together with a 
half-shift along $\wt S^1$\cite{9508064}. 
In this case if we consider an untwisted sector
state
that carries even unit of momentum along $\wt S^1$, then the
$(-1)^{F_L}$  for the surviving states must be 1, and as a result
when we evaluate $Tr(-1)^{F_L}$ over BPS 
supermultiplets of this type,
we simply get the total number of BPS states in this sector.
On the other hand for states
carrying odd units of winding along $\wt S^1$ the $(-1)^{F_L}$
quantum number must be $-1$, and as a result $Tr(-1)^{F_L}$ over such
BPS supermultiplets gives an answer that is negative
of the total number of BPS states in that sector.
Thus $\Omega_4$ is non-zero for both
sectors. Black hole entropy for this system will continue to vanish however
due to an argument of \cite{9712150}
that shows that the leading term in the
expression for black hole entropy has
a universal form independent of compactification and hence
vanishing of the
entropy of the black hole representing fundamental
string in type II string
theory on $T^6$ automatically implies the 
vanishing of the black hole 
entropy in this new asymmetric orbifold. 
Thus we see that there is a
mismatch between $\Omega_4$ and $e^{S_{BH}}$ 
even at the leading order.

The exact interpretation of this discrepancy is not completely clear to us.
We note however that unlike the heterotic case, where for a given charge $Q$ 
the total number of BPS states and hence also the index $\Omega_4$ is a 
function of the combination $N=Q^2/2$ only up to exponentially suppressed
terms, here $\Omega_4$ can take different values for different
charges even if $Q^2$ is the same for these charges.
Up to exponentially suppressed corrections, these 
values differ from each other by a $-$ sign. Thus if we replace $d_N$
by $(\Omega_{4})_{N}$ in eq.\refb{e2a},
we no longer have a well defined expression; the result depends on the
representative state that we use to compute $(\Omega_4)_{N}$. This
could be the reason why the correspondence
between black hole entropy and
$\Omega_4$ of BPS states does not hold in this case.

{\bf Acknowledgement}: I wish to thank  A.~Dabholkar, R.~Gopakumar,
D.~Jatkar, S.~Parvizi, J.~Samuel, D.~Surya Ramana, and A.~Tavanfar
for useful discussions.

\end{document}